\documentclass{IEEEtran}
\usepackage{mathrsfs}
\usepackage{amsfonts}
\usepackage{graphicx}   % extended graphics package
\usepackage{amsmath}    % math package
\usepackage{array}  % extended styles for tables
\usepackage{hhline}     % extended styles for tables
\usepackage{pslatex}    % so that we've got a clean pdf file afterwards
\usepackage{txfonts}
\usepackage{mathrsfs}
\usepackage{amssymb}
\usepackage{epsfig}
\usepackage{graphicx}
\usepackage{subfig}
\usepackage{setspace}
\usepackage{txfonts}                                                          % paper
\usepackage{float}

\makeatletter
\usepackage{mathrsfs}\usepackage{amsfonts}% extended graphics package
\usepackage{array}% extended styles for tables
\usepackage{hhline}% extended styles for tables
\usepackage{pslatex}% so that we've got a clean pdf file afterwards
\usepackage{txfonts}\usepackage{mathrsfs}%\usepackage{times}
\usepackage{epsfig}\usepackage{setspace}\usepackage{txfonts}% paper
\usepackage{float}\IEEEoverridecommandlockouts                              % This command is only
\newcommand{\sgn}{\mathop{\mathrm{sgn}}}

\title{\LARGE SymFET: A Proposed Symmetric Graphene Tunneling Field Effect Transistor}
\author{Pei Zhao, Randall M. Feenstra, Gong Gu and Debdeep Jena% <-this % stops a space

\thanks{\small P. Zhao and D. Jena are with the Department of Electrical Engineering, University
of Notre Dame, Notre Dame, IN 46556 USA,
        {\texttt{\small djena@nd.edu}}.

\small R. M. Feenstra is with Dept. Physics, Carnegie Mellon
University, Pittsburgh, PA 15213, USA.

\small G. Gu is with Dept. Electrical Engineering and Computer
Science, University of Tennessee, Knoxville, TN 37996,USA.
         }

         }

\@ifundefined{showcaptionsetup}{}{%
 \PassOptionsToPackage{caption=false}{subfig}}
\usepackage{subfig}
\makeatother

\begin{document}
\maketitle %\thispagestyle{empty} \pagestyle{empty}
\begin{abstract}
In this work, an analytical model to calculate the channel potential
and current-voltage characteristics in a Symmetric tunneling
Field-Effect-Transistor (SymFET) is presented.  The current in a
SymFET flows by tunneling from an n-type graphene layer to a p-type
graphene layer.  A large current peak occurs when the Dirac points
are aligned at a particular drain-to-source bias $V_{DS}$. Our model
shows that the current of the SymFET is very weakly dependent on
temperature. The resonant current peak is controlled by chemical
doping and applied gate bias. The on/off ratio increases with
graphene coherence length and doping. The symmetric resonant peak is
a good candidate for high-speed analog applications, and can enable
digital logic similar to the BiSFET. Our analytical model also
offers the benefit of permitting simple analysis of features such as
the full-width-at-half-maximum (FWHM) of the resonant peak and
higher order harmonics of the nonlinear current.  The SymFET takes
advantage of the perfect symmetry of the bandstructure of 2D
graphene, a feature that is not present in conventional
semiconductors.

\end{abstract}

\section{Introduction}

Graphene is an atomically thin two dimensional (2D) crystal
\cite{Geim_07}.  Due to the high mobility of carriers in it, their
linear dispersion and perfect 2D confinement, graphene is being
considered as a channel material for future electronic devices.
However, some challenges such as opening a finite bandgap for
digital applications still remain.

A distinguishing feature of graphene is its symmetric electronic
bandstructure.  The valence band is a perfect mirror image of the
conduction band about the Dirac point.  Such symmetry carries over
to gapped 2D crystals such as hexagonal Boron Nitride (h-BN), and
less so to transition metal dichalcogenides (such as Molybdenum
Disulfide, MoS$_2$)\cite{Neto_11}. This unique bandstructure {\em
symmetry} has not been sufficiently exploited for active device
applications till date.

Most studies of graphene based devices have focused on carrier
transport in the 2D plane of the crystal.  Recently, however,
carrier transport {\em out of the plane}, i.e., vertical to the
graphene sheet, has received increased attention.  These studies of
out-of-plane charge transport in 2D crystals have been motivated by
the proposal of the bilayer pseudo-spin FET (BiSFET) in 2009
\cite{Banerjee_09}.  The BiSFET exploits the fact that two graphene
layers can be placed in close proximity and if populated by
electrons and holes, the strong Coulomb attraction between them can
lead to exciton formation. Excitons are bosonic quasiparticles, and
can undergo condensation below a certain critical temperature. Since
the Fermi degeneracy in graphene is tunable over a large energy
window, the critical temperature for the excitonic condensate has
been calculated to be higher than room temperature. The formation of
the excitonic condensate is expected to lead to a macroscopic
tunneling current between the layers.  Similar behavior has been
observed at low temperatures and at high magnetic fields in coupled
AlGaAs/GaAs electron-hole bilayers
\cite{Eisenstein_91},\cite{Brown_94}.  The BiSFET thus has the
potential to realize many-body excitonic tunneling phenomena at room
temperature.  The power dissipation in computating using the
functionality of BiSFET is predicted to be many orders lower than
conventional CMOS switching.

Stacking of different 2D crystals leads to a novel class of
heterostructures \cite{Dean_10}. For example, stacking graphene with
BN results in a smooth surface, since BN shares the same hexagonal
lattice structure with graphene.  The absence of out-of-plane
covalent bonds implies that strain effects in similar lattice
mismatched heterostructures based on sp$^3$-bonded 3D crystals are
much reduced, or perhaps even eliminated.  At low carrier
concentration $n \thicksim$ 10$^{11}$ cm$^{-2}$, the device exhibits
mobilities of the order of 100,000 cm$^{2}$/V.s at room temperature,
which is much higher than for graphene on SiO$_{2}$ or SiC
\cite{Mayorov_11}. Electron transport {\em out} of the plane of
graphene has also started receiving attention in experiments. For
example, a graphene/BN/graphene sandwich heterostructure was
recently reported \cite{Britnell_12}, and interlayer electron
transport was measured. This stacked graphene/BN/graphene
heterostructure showed a room temperature switching ratio of 50 and
10,000 for a similar graphene/MoS$_2$/graphene structure
\cite{Britnell_12}. Another recent report \cite{Yang_12} showed that
electrons can be moved from 2D graphene into and out of Silicon to
form a variable barrier-height device (the Barristor).

As charge transport out of the plane of graphene receives increasing
attention, a pertinent question emerges upon careful analysis of the
proposed BiSFET device. What is the expected behavior of a similar
device structure consisting of a graphene-insulator-graphene (GIG)
p-n junction heterostructure, but in the {\em absence} of the
many-body excitonic condensate?  Similar devices already exist in
III-V resonant tunneling diodes, where single-particle tunneling
itself leads to a number of interesting and useful quantum phenomena
that persist at room temperature.  Negative differential resistance
is one such effect.  Single-particle tunneling current transport has
found enhanced attention recently in homojunction and heterojunction
Tunneling Field-Effect Transistors (TFETs).  It has been measured
across various semiconductor heterostructures at room temperature,
highlighting its robustness \cite{Seabaugh_10}.

\begin{figure*}[t]
\begin{centering}
\subfloat[]{\includegraphics[width=0.31\textwidth]{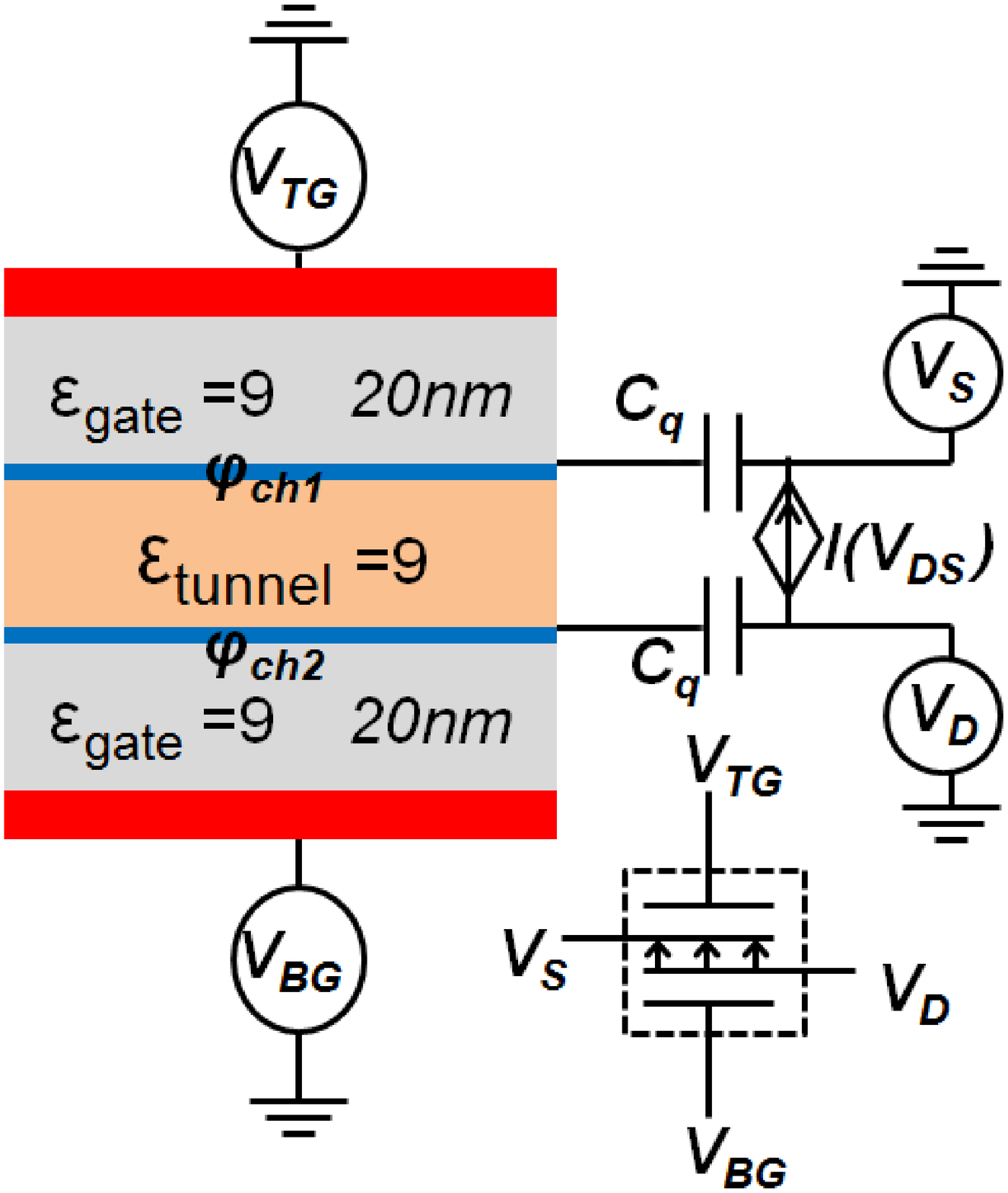}}
\subfloat[]{\includegraphics[width=0.48\textwidth]{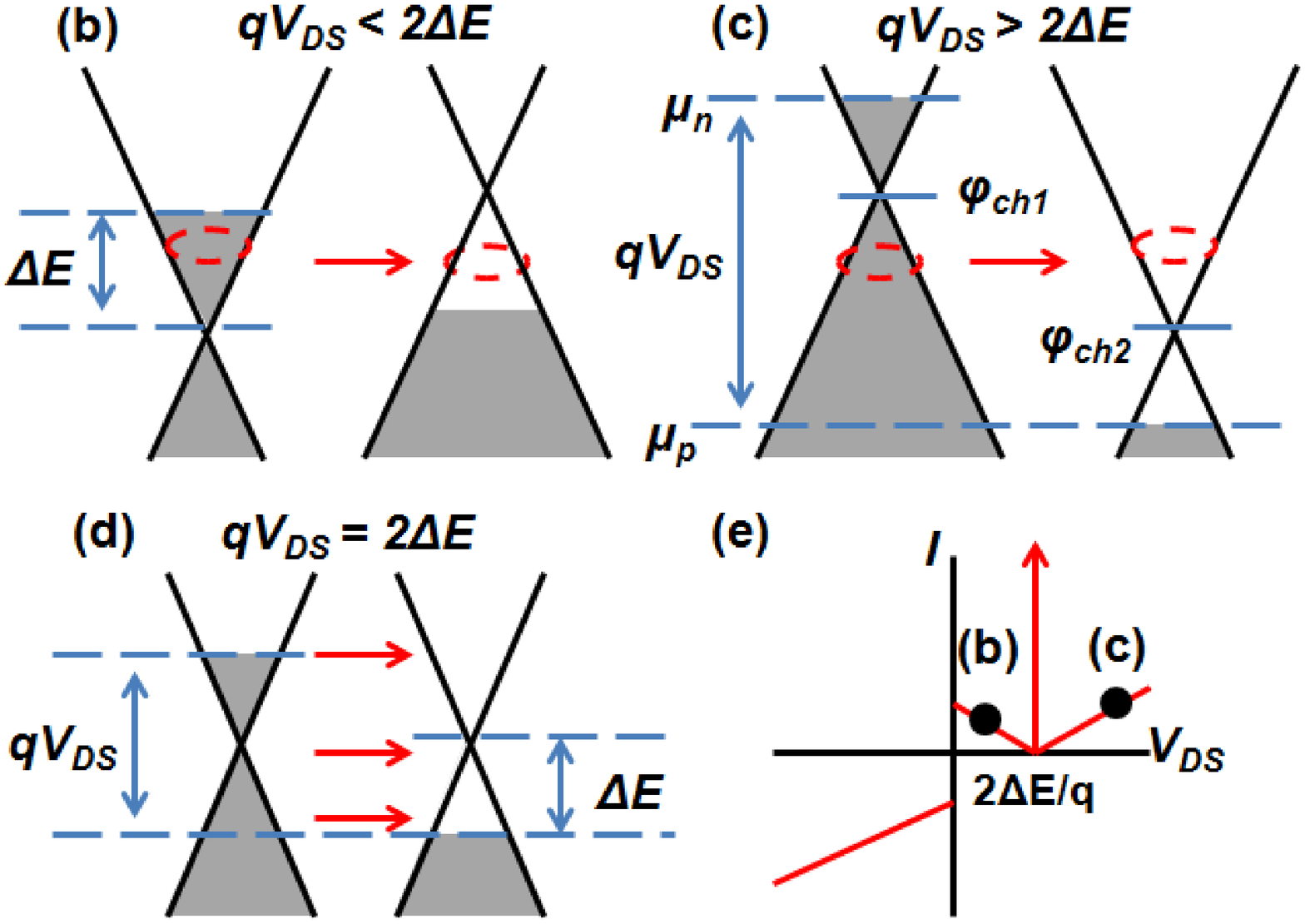}}
\par\end{centering}

\centering{}\caption{ (a) Sketch of the SymFET and the energy-band
diagrams for a doped graphene/insulator/graphene junction, at
voltages of (b) $qV_{DS}<2\Delta E$, (c) $qV_{DS}>2\Delta E$, and
(d) $qV_{DS}=2\Delta E$. A qualitative current-voltage $I-V_{DS}$
characteristic is shown in (e). The inset in (a) shows the symbol
defined for the SymFET.} \label{fig:fig1}
\end{figure*}

One major novel feature of graphene is the perfect symmetry of the
bandstructure, which can lead to enhanced functionality.  Motivated
by the above question, we recently calculated the single-particle
interlayer tunneling current-voltage curves explicitly for finite
area two-terminal GIG heterostructures \cite{Freenstra_12}.  The
general finding was that at most interlayer bias voltages, energy
and momentum conservations force a small tunneling current to flow
at one particular energy halfway between the two Dirac points.
However, at a particular interlayer voltage when the Dirac points of
the p- and n-type graphene layers align, a very large interlayer
tunneling current flows. This is because energy and momentum are
conserved in this process for all electron energies between the
quasi-Fermi levels of the n- and p-type graphene layers.  The $I-V$
curve is dominated by a Dirac-delta function-like peak at the
critical interlayer voltage, and smaller currents at all other
voltages.  Our explicit calculation of the tunneling current also
showed that the effect is highly robust to temperature, but less
robust to rotational misalignment of the two graphene layers. This
surprising, yet conceptually simple behavior of the 2-terminal GIG
device leads naturally to the question: how will a transistor
geometry with the single-particle GIG tunneling junction as its
channel behave?

In this work, we extend the detailed physical model of the 2-terminal GIG device described
in \cite{Freenstra_12} to a {\em Sym}metric graphene tunneling {\em F}ield {\em E}ffect {\em T}ransistor,
which we call the ``SymFET'' since its unique characteristics derive from the symmetry of
the bandstructure.  We derive analytical expressions for the channel potential and current-voltage
characteristics of the SymFET.  Possible logic and high frequency applications are also discussed.

\section{Device Model}

We assume a symmetric device structure as shown in
Fig.~\ref{fig:fig1}(a). Two graphene layers are separated by an
insulator, and this GIG structure is sandwiched between a top and
bottom gate.  Ohmic contacts are formed to the two graphene layers
individually representing the source (S) and the drain (D). The top
and bottom gate voltages $V_{TG}, V_{BG}$ control the quasi-Fermi
levels $\mu_{n}$ and $\mu_{p}$ in the top and bottom layers of
graphene. The gate insulator thicknesses of both gates are assumed
to be the same. The quasi-Fermi level is $\Delta E$ above the Dirac
point in the n-type graphene layer and below the Dirac point in the
p-type graphene layer.  This is indicated in Fig.~\ref{fig:fig1}(b),
(c), and (d). The top and back gates are symmetric $V_{TG}=-V_{BG}$,
and the drain-source voltage is $V_{DS}=V_{D}-V_{S}$. The insert in
Fig.~\ref{fig:fig1} (a) shows a proposed device symbol for the
SymFET.

As shown in Fig.~\ref{fig:fig1}(b) and (c), under S/D biases when
the two Dirac points are misaligned, only a single energy (and
lateral k-momentum ring) in the Dirac cone meets the requirement of
simultaneous energy and momentum conservation, and thus the
tunneling current is small.  At $V_{DS} = 2 \Delta E / q$, however,
the two Dirac points align, and electrons at {\em all} energies
between the quasi-Fermi levels satisfy energy and momentum
conservation.  A large tunneling current is thus expected; a
resonant current peak originating from the perfectly symmetric
bandstructure of the graphene layers should result.  Since graphene
is not a metal, a part of the applied voltage will drop in the
graphene layer itself. This effect of the finite density of states
is captured in the quantum capacitance of graphene and is included
in the model. This is critical since the gate capacitance $C_{g} =
\epsilon_{g}/t_{g}$ and the tunneling capacitance $C_{t} =
\epsilon_{t}/t_{t}$ are large and can reach the quantum capacitance
limit easily \cite{Luryi_88}.  For simplicity of the calculation we
use the $T \rightarrow 0K$ limit for quantum capacitance:

\begin{equation} \label{eq:Cq}
C_{q}=\frac{2}{\pi}\frac{|\Delta E|}{(\hbar v_{F}/q)^{2}},
\end{equation}
where $q$ is the single electron charge, $v_F$ is the Fermi velocity
in graphene, $\hbar$ is the reduced Planck constant. It can be
verified that this approximation is a good one at room temperature.

The source and drain electrodes are assumed to be perfect ohmic
contacts for simplicity.  In practice, the contact resistance will
also force a voltage drop, and can be added on top of the intrinsic
model.  The interlayer tunneling current calculated in
\cite{Freenstra_12} depends on the interlayer voltage difference.
Therefore, to find the behavior of the 4-terminal SymFET, potentials
of the two graphene layers $\varphi_{ch1}$ and $\varphi_{ch2}$ need
to be identified as a function of the gate and drain/source biases
(the channel potentials of graphene and Fermi levels are referenced
to the aligned channel potentials at flat band.). To do so, we
invoke the charge neutrality condition:

\begin{equation} \label{eq:CNC1}
(\frac{\varphi_{ch1}}{q}+V_{TG})C_{g}+(\frac{\varphi_{ch1}}{q}-\frac{\varphi_{ch2}}{q})C_{t}+(\frac{\varphi_{ch1}}{q}-\frac{\mu_{n}}{q})C_{q}/2+qN=0,
\end{equation}
\begin{equation} \label{eq:CNC2}
(\frac{\varphi_{ch2}}{q}+V_{BG})C_{g}+(\frac{\varphi_{ch2}}{q}-\frac{\varphi_{ch1}}{q})C_{t}+(\frac{\varphi_{ch2}}{q}-\frac{\mu_{p}}{q})C_{q}/2-qN=0,
\end{equation}
where $V_{TG}=-V_{BG}=V_G$, $N=\Delta E_{doping}^2/\pi (\hbar
v_{F})^{2}$ is the chemical doping concentration, and we assume the
work functions of the metals are matched with the undoped graphene
sheets which gives the ¡®flat-band¡¯ conditions at zero gate bias.
The factors 1/2 in the third terms of equations \eqref{eq:CNC1} and
\eqref{eq:CNC2} are due to the linear dependence on $\Delta E$ in
the graphene quantum capacitance, which is a differential
capacitance.

Then we take equation \eqref{eq:CNC1} minus equation
\eqref{eq:CNC2}, use relationships: $qV_{DS}=2\Delta E+
\varphi_{ch1}-\varphi_{ch2}$, $qV_{DS}=\mu_n- \mu_p$,
$\mu_n-\varphi_{ch1}=\Delta E$, and $\varphi_{ch2}-\mu_p=\Delta E$,
we can form a quadratic equation with the only unknown parameter
$\Delta E$:

\begin{equation} \label{eq:CNC3}
(V_{DS}-2 \frac{\Delta E}{q}+2V_{G})C_{g}+2(V_{DS}-2\frac{\Delta
E}{q})C_{t}-\frac{2q \Delta E^2}{\pi (\hbar v_{F})^{2}} +2qN=0,
%%\label{chargeneutrality}
\end{equation}
and the solution is

%\newcounter{MYtempeqncnt}
%\begin{figure*}[!t]
%\normalsize
%\setcounter{MYtempeqncnt}{\value{equation}}
%%\setcounter{equation}{1}
\begin{equation} \label{eq:Vch1}
\begin{aligned} \Delta E(V_{G},V_{DS})&= -\frac{(2C_{t}+C_{g})\pi(\hbar v_{F}/q)^{2}}{2}  \\
 & +\Bigg\{  \frac{ (2C_{t}+C_{g})^2 \pi^2(\hbar v_{F}/q)^{4} }{4}  \\
 & +\frac{\pi(\hbar v_{F})^{2}}{2q}[(V_{DS}+2V_G)C_g+2C_tV_{DS} +2qN ]\Bigg\} ^{\frac{1}{2}} ,
\end{aligned}
\end{equation}
%\setcounter{equation}{\value{MYtempeqncnt}} \hrulefill \vspace*{4pt}
%\end{figure*}
The electrostatic model used here is based on a 1D approximation,
ignoring the intra-graphene layer potential distribution and current
flows.  A more rigorous treatment requires the solution of the 2D
Poisson equation, which is suggested for future work.

The analytical expression for the interlayer tunneling current at
zero temperature was derived in \cite{Freenstra_12}. When the Dirac
points in the two graphene layers are misaligned, the nonresonant
tunneling current is:

\begin{equation} \label{eq:Id1}
I=G_{1}(\frac{2\Delta E}{q}-V_{DS}),\ (0<qV_{DS}<2\Delta E),
\end{equation}

\begin{equation} \label{eq:Id2}
I=G_{1}(V_{DS}-\frac{2\Delta E}{q}),\ (qV_{DS}>2\Delta E\ or\
qV_{DS}<0),
\end{equation}
where the prefactor conductance
$G_{1}=\frac{q^{2}A}{2\hbar}(\frac{\hbar\kappa u_{12}^{2}e^{-\kappa
t_t}}{mdv_{F}})^2$, $\kappa$ is a decay constant for the tunneling
current in barrier, $m$ is the free electron mass, $d$ is the
normalization constant for z-component wavefunction in graphene,
$u_{12}$ is a constant of order unity, and $A=L^{2}$, with $L$ being
the coherence length of graphene (size of ordered areas in graphene
film). In this work, we assume the graphene size is $A$ with
coherence in all area.

The resonant current is a perfect Dirac-delta function at
$qV_{DS}=2\Delta E$ for infinitely wide graphene sheets. For finite widths of length $L$, it is
broadened to \cite{Freenstra_12}:

\begin{equation} \label{eq:Id3}
I=\frac{1.6}{\sqrt{2\pi}}G_{1}\frac{L\Delta
E^{2}(2u_{11}^{4}+u_{12}^{4})}{u_{12}^{4}q\hbar
v_{F}}\exp[-\frac{A}{4\pi}(\frac{qV_{DS}-2\Delta E}{\hbar
v_{F}})^{2}],
\end{equation}
where $u_{11}$ is a constant of order unity, similar to $u_{12}$.
The total current is the combination of Equation \eqref{eq:Id1} or
Equation \eqref{eq:Id2} summed up together with Equation
\eqref{eq:Id3}.

Equations \eqref{eq:Id1} - \eqref{eq:Id3} were derived assuming zero
temperature. At finite temperature, to capture the thermal
occupation of states, the current needs to be calculated by
including the Fermi-Dirac distributions in the integral over all
states  \cite{Freenstra_12}. Equations \eqref{eq:IdRT1} and
\eqref{eq:IdRT2} are the finite temperature expressions
corresponding to Equations \eqref{eq:Id1} and \eqref{eq:Id2},
derived by a direct extension of the theory in our previous work
\cite{Freenstra_12},

\begin{subequations}\label{eq:IdRT1}
\begin{align}
%\begin{equation}
I&=G_{1}\frac{4\hbar^{2}v_{F}^{2}}{q}\int\limits_{0}^{+\infty} \Big\{ k[f(E_{n,k}-\mu_n,T)-f(E_{p,k}-\mu_p,T)] \nonumber\\
        & \times\delta(2\Delta E-qV_{DS}-2\hbar v_{F}k)\Big\} dk \label{eq:IdRT1a}\\
%\end{equation}
%\begin{equation}
&=G_{1}(\frac{2\Delta E}{q}-V_{DS})[f(-qV_{DS}/2,T)-f(qV_{DS}/2,T)]
\label{eq:IdRT1b}\\
%\end{equation}
%\begin{equation}
&=G_{1}(\frac{2\Delta E}{q}-V_{DS})\tanh(\frac{qV_{DS}}{4k_BT}),\
(qV_{DS}< 2\Delta E) \label{eq:IdRT1c}
%\end{equation}
\end{align}
\end{subequations}

\begin{subequations}\label{eq:IdRT2}
\begin{align}
%\begin{equation}
I&=G_{1}\frac{4\hbar^{2}v_{F}^{2}}{q}\int\limits_{0}^{+\infty} \Big\{k[f(E_{n,k}-\mu_n,T)-f(E_{p,k}-\mu_p,T)] \nonumber\\
& \times\delta(qV_{DS}-2\Delta E-2\hbar v_{F}k)\Big\} dk \label{eq:IdRT2a}\\
%\end{equation}
%\begin{equation}
&=G_{1}(V_{DS}-\frac{2\Delta E}{q})[f(-qV_{DS}/2,T)-f(qV_{DS}/2,T)]
\label{eq:IdRT2b}\\
%\end{equation}
%\begin{equation}
&=G_{1}(V_{DS}-\frac{2\Delta E}{q})\tanh(\frac{qV_{DS}}{4k_BT}),\
(qV_{DS}> 2\Delta E) \label{eq:IdRT2c}
%\end{equation}
\end{align}
\end{subequations}
where subscripts n and p refer to the n-type (top) and p-type
(bottom) electrodes, respectively, and with the Fermi-Dirac
distribution for arguments $\delta E$ and T given by $f(\delta
E,T)\equiv1/[1+\exp(\delta E/k_BT)]$. Following Ref.
\cite{Freenstra_12} we have for Eq. \eqref{eq:IdRT1a}, with
$0<qV_{DS}<2\Delta E$, that $E_{n,k}-\mu_n=\hbar v_Fk-\Delta E$ and
$E_{p,k}-\mu_p=-\hbar v_Fk+\Delta E$, hence yielding Eqs.
\eqref{eq:IdRT1b} and \eqref{eq:IdRT1c}. Those equations also hold
for $qV_{DS}<0$, with the difference between the Fermi-Dirac
occupation factors changing sign. For Eq. \eqref{eq:IdRT2a}, with
$qV_{DS}>2\Delta E$, we have $E_{n,k}-\mu_n=-\hbar v_Fk-\Delta E$
and $E_{p,k}-\mu_p=\hbar v_Fk+\Delta E$, yielding
Eqs.\eqref{eq:IdRT2b} and \eqref{eq:IdRT2c}.

\begin{figure}[t]
\begin{centering}
\subfloat[]{\includegraphics[width=0.25\textwidth]{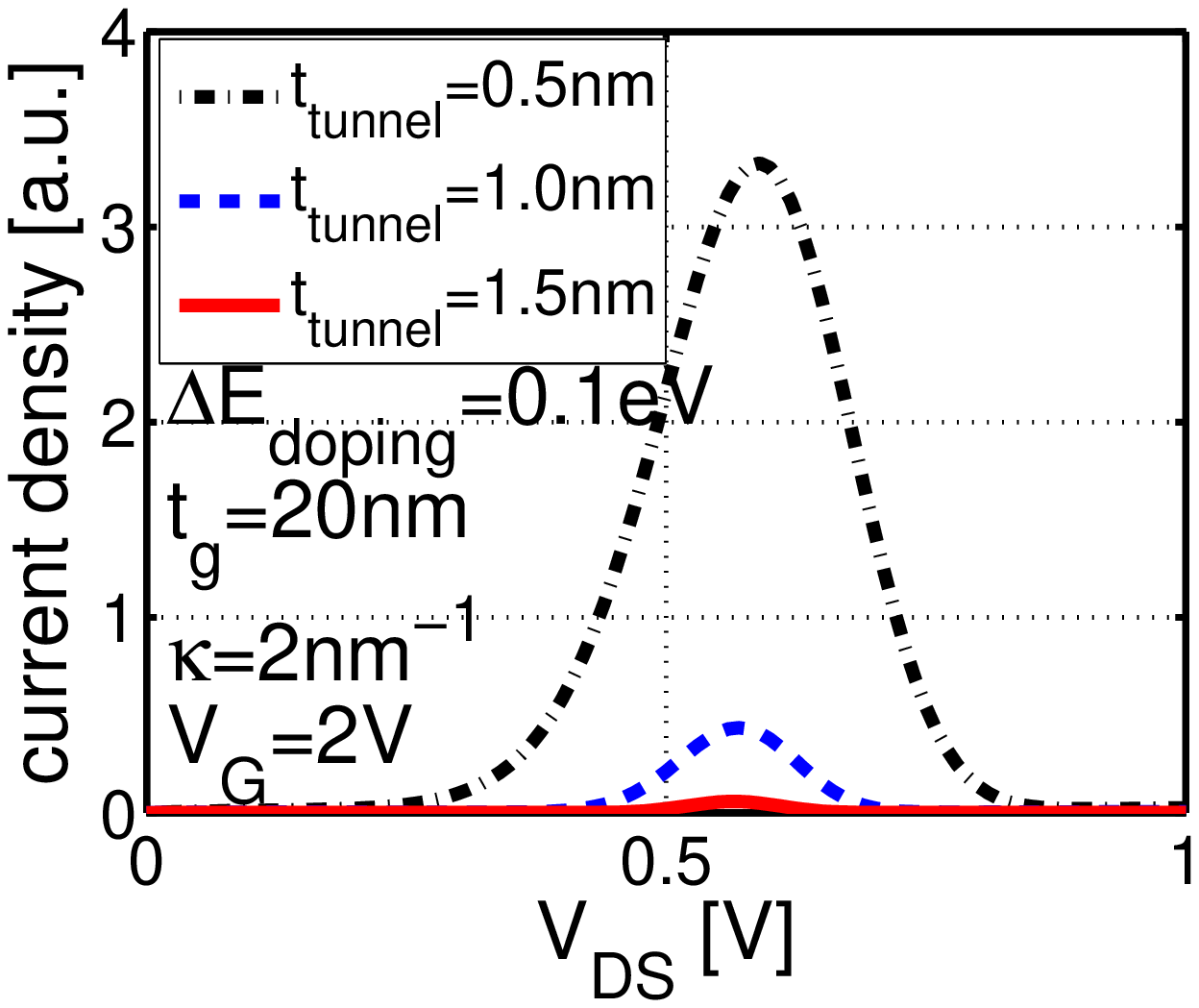}

}\subfloat[]{\includegraphics[width=0.25\textwidth]{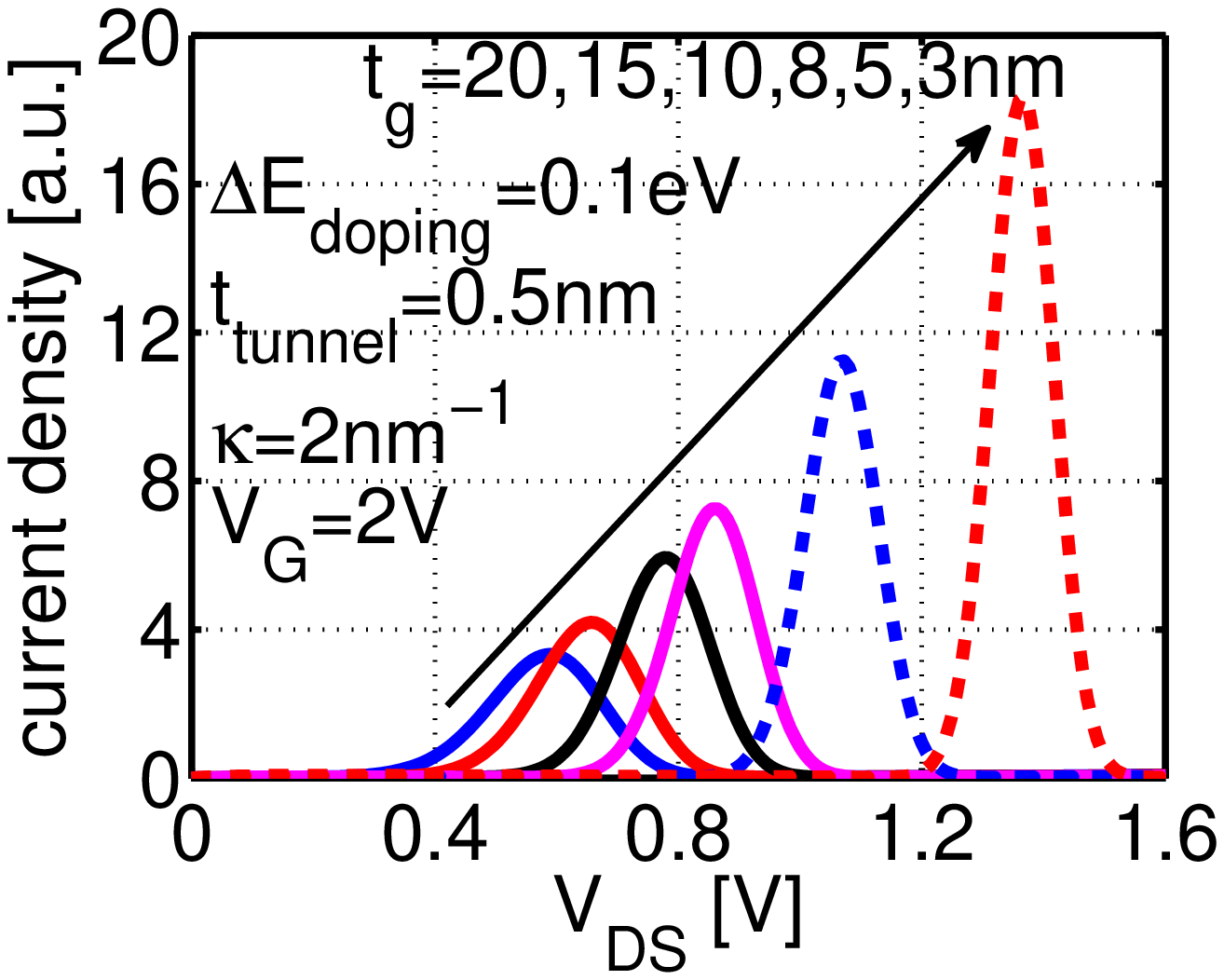}

}
\par\end{centering}

\centering{}\caption{$I_{D}$ vs.$V_{DS}$ characteristics showing
scaling with (a) the tunneling insulator thickness and (b) the gate
insulator thickness (an arbitrary $\kappa$ value is chosen here to
give a clear illustration of current density scaling).}
\label{fig:fig2}
\end{figure}

A finite-temperature correction to Eq. \eqref{eq:Id3} is more
complicated, since that equation itself is a significant
approximation\cite{Freenstra_12}. Nevertheless, the dominant term in
its temperature dependence can be recognized as the increased number
of states available for tunneling, in the resonant situation of Fig.
1(d) with $qV_{DS}=2\Delta E$. For this resonant situation the Dirac
points of the two electrodes are aligned, at energy $E_d$. This
number of states is given simply by

\begin{equation} \label{eq:IdRT3a}
N_s(T)=\int\limits_{-\infty}^{+\infty}\rho(E-E_d)[f(E-E_d-\Delta
E,T)-f(E-E_d+\Delta E,T)]dE,
\end{equation}
where $\rho(E)=2|E|/[\pi(\hbar v_F)^2]$ is the density-of-states
(DOS) per unit area of graphene. Our correction to Eq.
\eqref{eq:Id3} is then made by multiplying it by $N_s(T)/N_s(0)$,
with $N_s(0)=2\Delta E^2/[\pi(\hbar v_F)^2]$. Expressing the result
in terms of Fermi-Dirac integrals of order 1, $\Im_1$, we have

\begin{equation} \label{eq:ratio}
\frac{N_s(T)}{N_s(0)}=\frac{2(k_BT)^2}{\Delta
E^2}[\Im_1(\frac{\Delta E} {k_BT})-\Im_1(-\frac{\Delta E} {k_BT})]
\end{equation}
where $\Im_1(x)=\int_{0}^{+\infty} t/[1+\exp(t-x)]dt$.

Equations \eqref{eq:IdRT1} and \eqref{eq:IdRT2}, as well as the
product of Eqs. \eqref{eq:Id3} and \eqref{eq:ratio}, are valid only
in the limit of large $L$, although the $L$ value at which the
approximations break down is different for the two cases. In the
former case, the dominant term in the slope of $I(V_{DS})$ at
$V_{DS}=0$ is $qV_{DS}/4k_BT$, arising from the tanh terms, and this
slope becomes very large for small $T$. However, the exact solution
for the current \cite{Freenstra_12} yields a slope that is limited
by the $L$ value. We find that simply multiplying Eqs.
\eqref{eq:IdRT1} and \eqref{eq:IdRT2} by a factor of
$\tanh(LqV_{DS}/\pi \hbar v_F)$ yields a slope at that agrees very
well with the exact solution, and it does not significantly affect
the $I(V_{DS})$ curve elsewhere. For the product of Eqs.
\eqref{eq:Id3} and \eqref{eq:ratio}, the current from Eq.
\eqref{eq:Id3} is nonzero at $V_{DS}$=0 , with this discrepancy
being significant only for sufficiently small $L$ values. In this
case we find that multiplying the product of Eqs. \eqref{eq:Id3} and
\eqref{eq:ratio} by a factor of $\tanh (LqV_{DS}/2\pi \hbar v_F)$
solves this problem of the nonzero current at $V_{DS}=0$ , and it
also produces a slope at $V_{DS}=0$ that agrees fairly well with the
exact solution. Hence, our final formula for the total current is
given by

\begin{equation} \label{eq:IdRT4a}
\begin{aligned}  I&= G_1(V_{DS}-\frac{2\Delta E}{q}) \sgn(V_{DS}-\frac{2\Delta E}{q})\tanh(\frac{qV_{DS}}{4k_BT})\tanh(\frac{LqV_{DS}}{\pi \hbar v_F}) \\
 &+ \Bigg\{ \frac{1.6}{\sqrt{2\pi}}G_{1}\frac{L\Delta
E^{2}(2u_{11}^{4}+u_{12}^{4})}{u_{12}^{4}q\hbar
v_{F}}\exp[-\frac{A}{4\pi}(\frac{qV_{DS}-2\Delta E}{\hbar
v_{F}})^{2}]\\
& \times\frac{N_s(T)}{N_s(0)}\tanh(\frac{LqV_{DS}}{2\pi \hbar v_F})
\Bigg\}
\end{aligned}
\end{equation}
where sgn$(V_{DS}-2\Delta E/q)$ equals 1 for $V_{DS}> 2\Delta E/q$,
0 for $V_{DS}= 2\Delta E/q$ , or -1 for $V_{DS} < 2\Delta E/q$.

The single particle tunneling model used in this work captures the
relevant physics e.g. wave function overlap (detailed derivation in
Ref. \cite{Freenstra_12}), even in the case of strong inter-layer
interactions at small insulator thicknesses. The only approximation
we made is that the electronic structure (band structure) of the GIG
system is the same as two non-interacting graphene sheets. We have
justified the approximation as follows: any modification to the band
structure due to the interaction will be near the Dirac energy. At
sufficiently high doping $\Delta E$, the error is negligible.

\begin{figure}[t]
\begin{centering}
\subfloat[]{\includegraphics[width=0.24\textwidth]{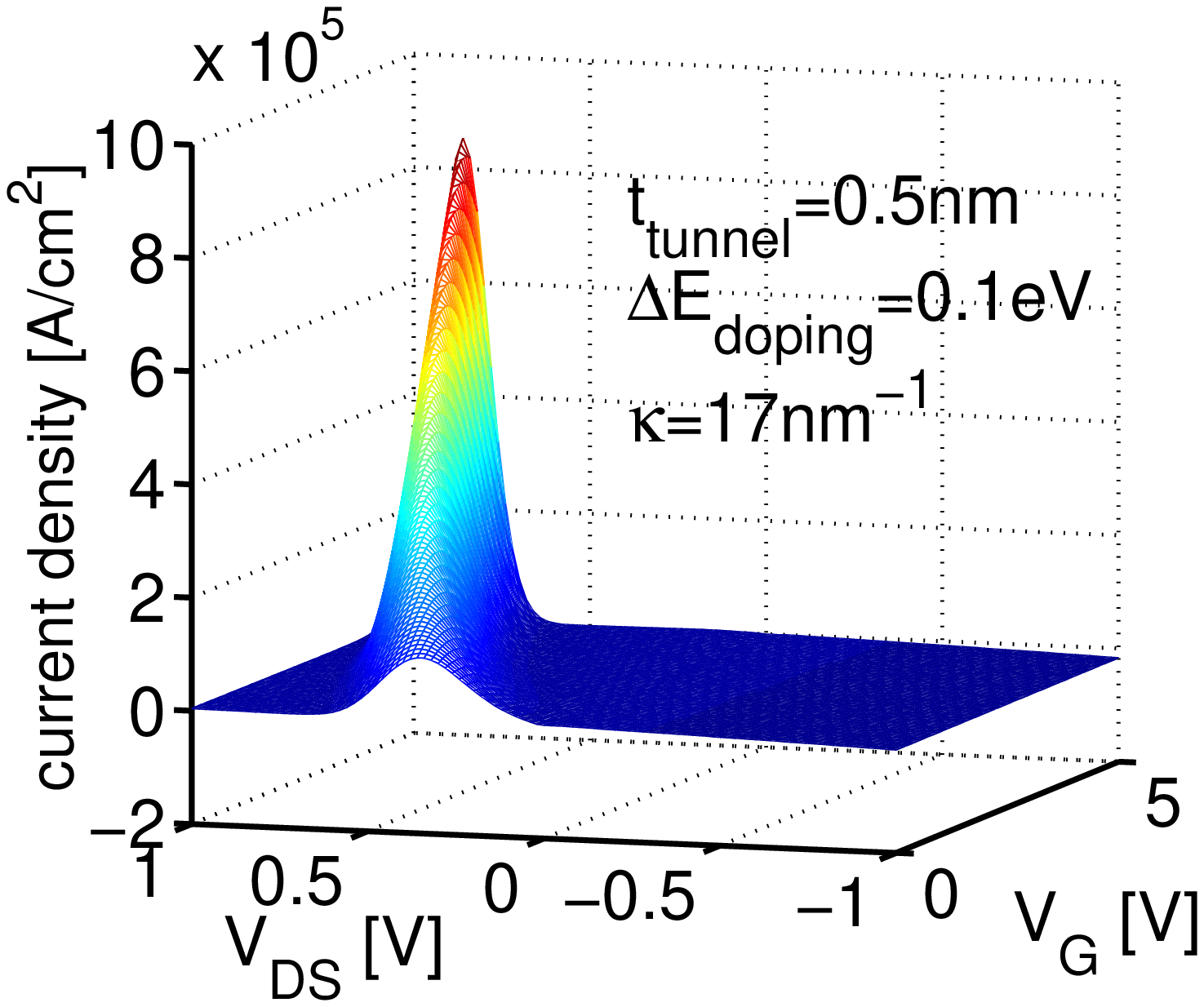}

}\subfloat[]{\includegraphics[width=0.24\textwidth]{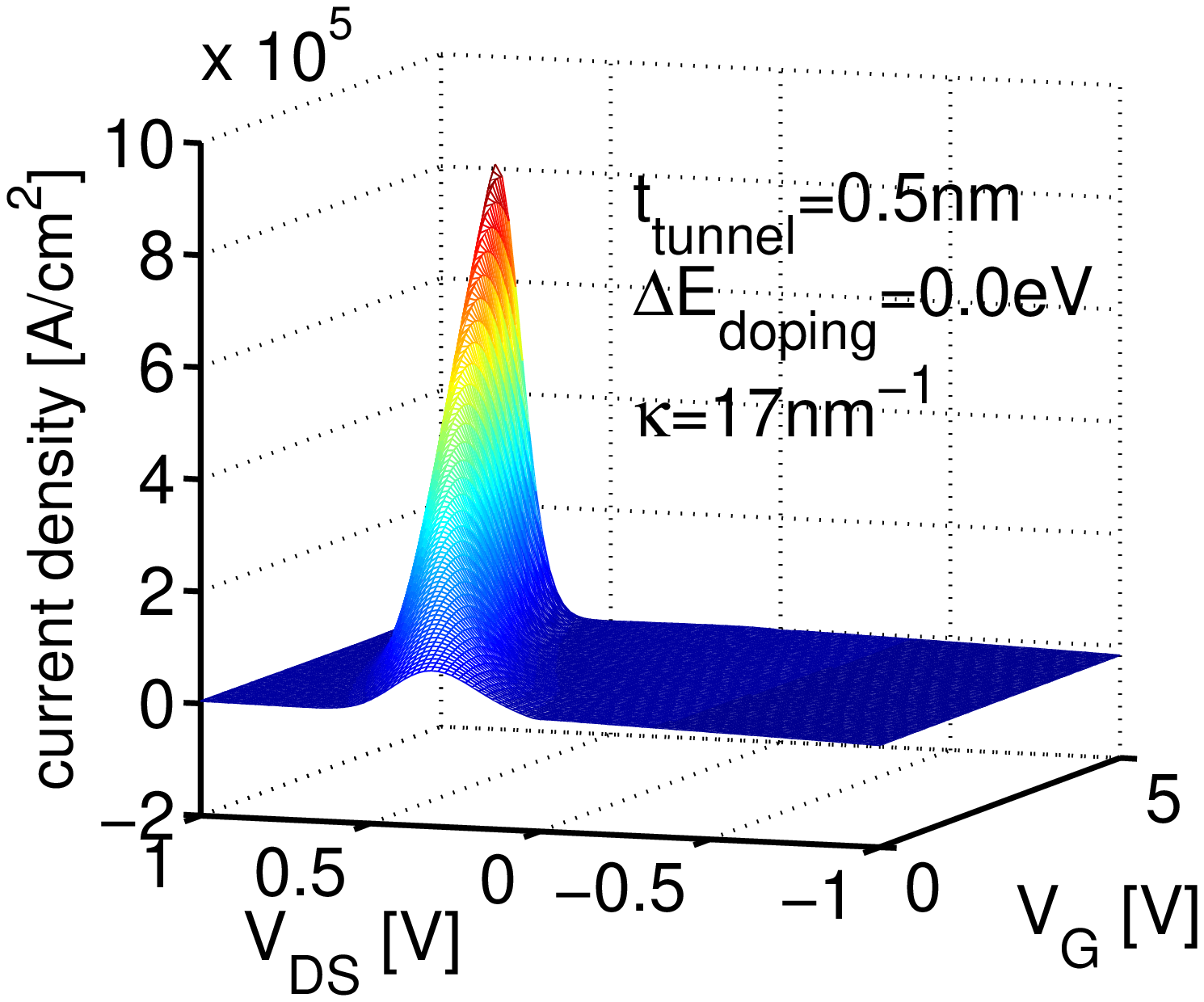}

}
\par\end{centering}

\centering{}\caption{The contour plot of the complete bias space for
(a) chemical doping $\Delta E_{doping}=0.1eV$ and (b) no chemical
doping. } \label{fig:fig3}
\end{figure}

\section{Results and Discussions}

We present typical $I-V$ characteristics of the SymFET
 at room temperature first. The comparison of $T=300$ K and $T=0$ K
is discussed later. The values of the decay constant $\kappa$ can be
calculated from the complex band structure inside the bandgap of the
insulator based within the effective mass approximation
\cite{Britnell_12}.  Here, we use an estimated value $\kappa=17$
nm$^{-1}$, following footnote 14 of \cite{Freenstra_12}. The
chemical doping level is set to be $\Delta E_{doping}=0.1$ eV. A
finite coherence length $L=100$ nm is assumed; the effect of this
parameter on the on/off ratio for the device is discussed below.
When the SymFET scales down to 50nm or less, momentum conservation
does not scale well with device size. Scaling limits of SymFET will
be a subject of future study; the quantum confinement and quantized
transverse momentum in graphene nanoribbon also need to be
considered near scaling limits, as discussed in Ref. \cite{Masum
Habib_11}.

The resultant $I_{D}-V_{DS}$ characteristics with varying insulator
thickness are shown in Fig.~\ref{fig:fig2}.  The tunneling insulator
thickness $t_t$ is similar as the tunneling barrier thickness in
double quantum well heterostrucures \cite{Brown_94}. As
 $t_{t}$ increases the resonant peak current decreases, as expected. The
gate insulator can be a high-k material similar to that employed in
Si CMOS technology. In addition, 2D materials such as BN might be a
better choice to reduce the interface trap density since the
dangling bond can be reduced. The measured breakdown field is as
high as 7.94 MV/cm for BN \cite{Lee_11}. Thinner $t_{g}$ offers
better gate control and higher gate induced doping. When $t_{g}$
decreases, $\Delta E$ becomes larger at same gate bias. The resonant
peak moves to a higher bias and the peak current increases. For the
simulation results shown next, we fix the gate capacitance with
$t_{g}=20$ nm and dielectric contact $\epsilon_{g}=9$, and the
tunneling insulator thickness $t_{t}=0.5$ nm, and dielectric contact
$\epsilon_{t}=9$. (BN might have an even lower dielectric constant
$\epsilon_{BN}=3.5$ \cite{Dean_11}).

\begin{figure}[t]
\begin{centering}
\subfloat[]{\includegraphics[width=0.25\textwidth]{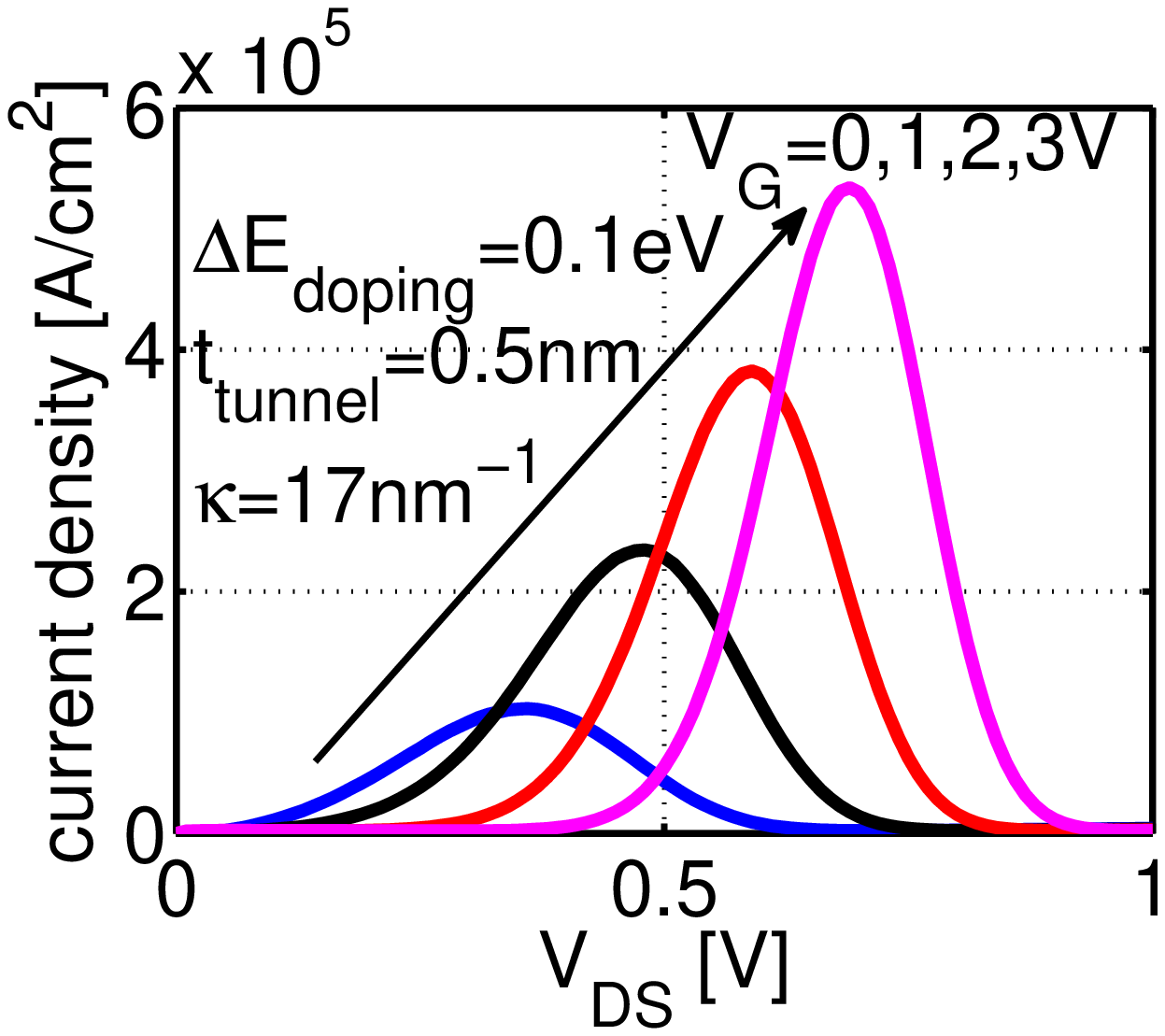}

}\subfloat[]{\includegraphics[width=0.25\textwidth]{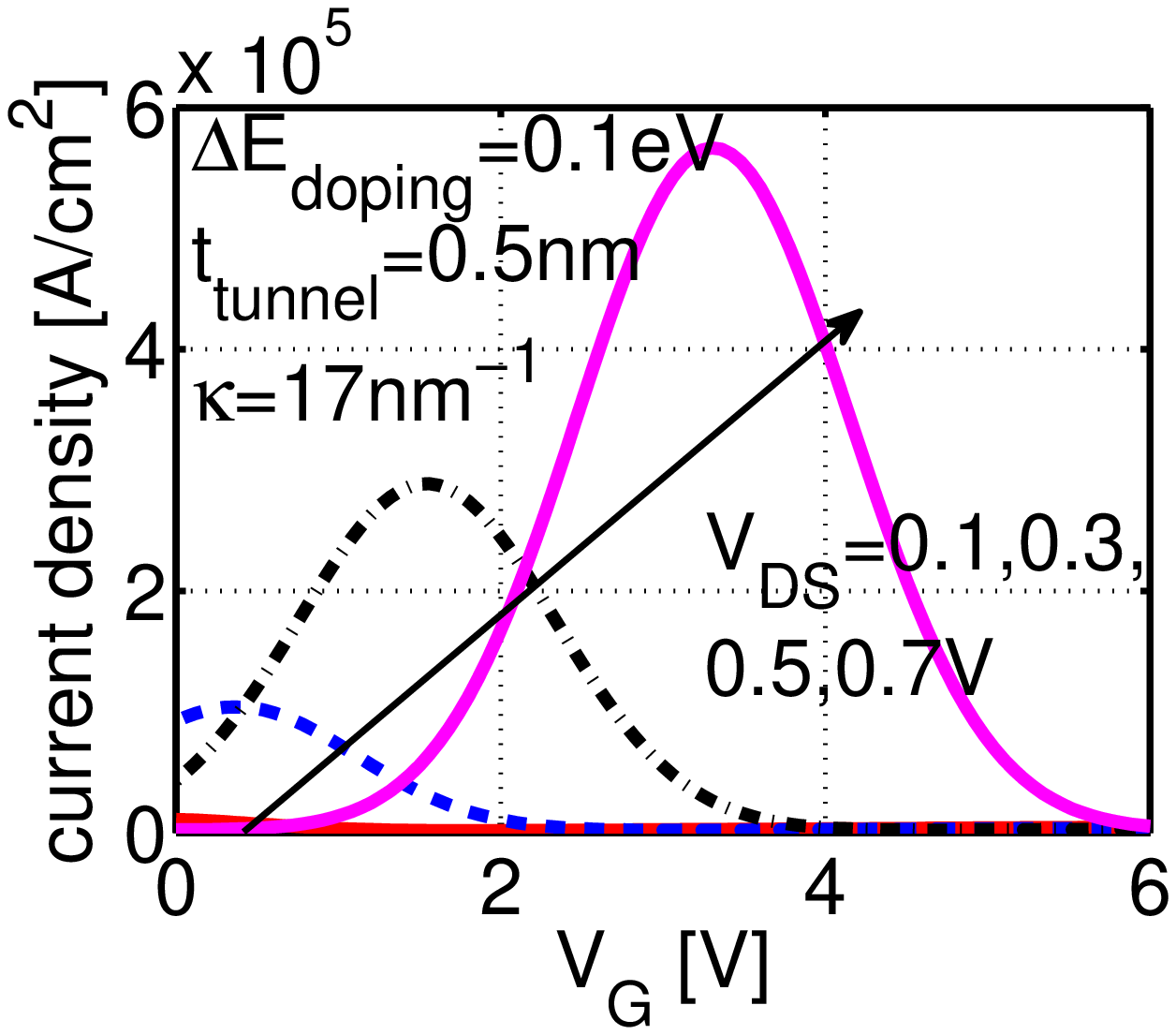}

}
\par\end{centering}

\centering{}\caption{ (a) $I_{D}$ vs. $V_{DS}$ curves with different
$V_{G}$, and (b) $I_{D}$ vs. $V_{G}$ at different $V_{DS}$. }
\label{fig:fig4}
\end{figure}

In Fig.~\ref{fig:fig3}, the entire bias phase space of the $I-V$
characteristics are shown. When the graphene sheets are chemically
doped (i.e., $\Delta E_{doping} = 0.1$ eV), $V_{DS}$ drives the
carriers tunneling between p- and n-type of graphene and the
resonant peak exists for $V_G$ = 0 V . With non-zero $V_G$, gate
electrostatic doping further increase $\Delta E $ and resonant peak
shift to higher $V_{DS}$. But the current is quite small in the
non-resonant region.

If we define the peak current as the $I_{on}$ and the current close
to $V_{DS} \sim 0$ as $I_{off}$, from Equation \eqref{eq:Id1} and
\eqref{eq:Id3} the effective on/off ratio is:

\begin{equation} \label{eq:onoff}
\frac{I_{on}}{I_{off}}=\frac{0.8}{\sqrt{2\pi}}\frac{L\Delta E}{\hbar
v_{F}}.
\end{equation}

The effective on/off ratio is independent of temperature (ignoring
the slight difference due to the Fermi tail), and increases with the
doping $\Delta E$ and the graphene size $L$. For $L=100$ nm the
ratio is $\sim$100 and $\sim$1000 for $L=1$ $\mu $m. We also point
out that, the on/off ratio is not necessarily a figure of merit for
the device (since the device might be employed for analog
applications where high modulation is not required). In a similar
vein, the resonant peak is symmetric in voltage, and represents a
rather strong negative differential resistance.  The
peak-to-valley-current-ratio (PVCR) of this NDR is identical to the
on/off ratio defined above.

In Fig.~\ref{fig:fig3} (b), we assume that $\Delta E_{doping}=0$ eV.
At $V_G=0$ V, $\Delta E$ is non-zero since drain bias also induces
doping, similar as the case of GIG junction in \cite{Freenstra_12}.
However, $\Delta E$ is small and the resonant peak is small at low
$V_{G}$. When $V_{G}$ increases, electrostatic doping induces an
appreciable resonant current, which further increases at higher
$V_{G}$.

The $I_{D}-V_{DS}$ characteristics at fixed $V_{G}$ is shown in
Fig.~\ref{fig:fig4} (a). The resonant behavior shows clear on and
off states without a saturation region.  Because we assume a
chemical doping of graphene, the SymFET with a resonant current peak
can operate at $V_{G}=0V$.  As mentioned above, the gate will induce
electrostatic doping in the graphene layer.  With larger $V_{G}$,
$\Delta E$ increases, the resonant condition $qV_{DS}=2\Delta E$
occurs at larger drain bias, and the resonant current peak moves to
the right. Higher $V_{G}$ induces more doping and thus large
on-state current. In the 2-terminal GIG device, the resonant current
peak is proportional to the coherence length $L$, and the width is
proportional to $1/L$ \cite{Freenstra_12}.  In the gated SymFET,
since the gate bias electrostatically dopes the graphene, it offers
the additional flexibility to adjust the on and off states.  In
Fig.~\ref{fig:fig4} (b), $I_{D}-V_{G}$ curves are shown with a
strong non-linear and resonant behavior, but with wider peaks.  When
$V_{G}$ is small and outside the resonant peak, the transconductance
is small, but is large in the peak condition.

Different from the sub-threshold region of MOSFETs, we refer to the
`off' state away from the resonant peak in the SymFET as the
non-resonant region. Although the graphene sheets are doped,
according to conservation laws, the current near $V_G=0$ V is low
($V_{DS}=0.7$ V, solid line in Fig.~\ref{fig:fig4} (b)). The gate
voltage modulates the doping potential $\Delta E$ in SymFET. By Eq.
\eqref{eq:Vch1}, at fixed $V_{DS}$, doping potential $\Delta E$ is a
sub-linear function of $V_G$. The current $I_{DS}(V_G)$ at
non-resonant region roughly follows the same hyperbolic form as the
first term of $I_{DS}(V_{DS})$ in Eq. \eqref{eq:IdRT4a}:

\begin{equation} \label{eq:IdVg}
\begin{aligned}  I\propto \tanh(\frac{qV_{G}}{4k_BT})\tanh(\frac{LqV_{G}}{\pi \hbar
v_F}).
\end{aligned}
\end{equation}
Hyperbolic functions may not offer a sharp subthreshold-swing (SS),
but the SymFET is more attractive for analog applications, where the
steep SS is not necessary.

The resonant current peak follows a normal distribution function
$I\sim \exp(- \frac {(qV_{DS}-\Delta E)^2 }{2 \sigma ^2 } )$, with
Full-width-at-half-maximum, $FWHM = 2.3548\sigma$, where $\sigma=
\frac{\sqrt{2 \pi} \hbar v_{F} }{L}  $ for the resonant peak in
SymFET. In fact, $\Delta E$ is not a constant but also dependent on
$V_{DS}$ and $V_{G}$, thus the FWHM shown here is only an
approximation. The reason why a narrower FWHM occurs for $V_{DS}$
compared with $V_G$ in Fig.~\ref{fig:fig4} relates with the large
gate insulator thickness (weaker gate control). Smaller FWHM of
$V_{G}$ can be achieved with thinner gate insulator (not shown
here).

\begin{figure}[t]
\begin{centering}
\subfloat[]{\includegraphics[width=0.25\textwidth]{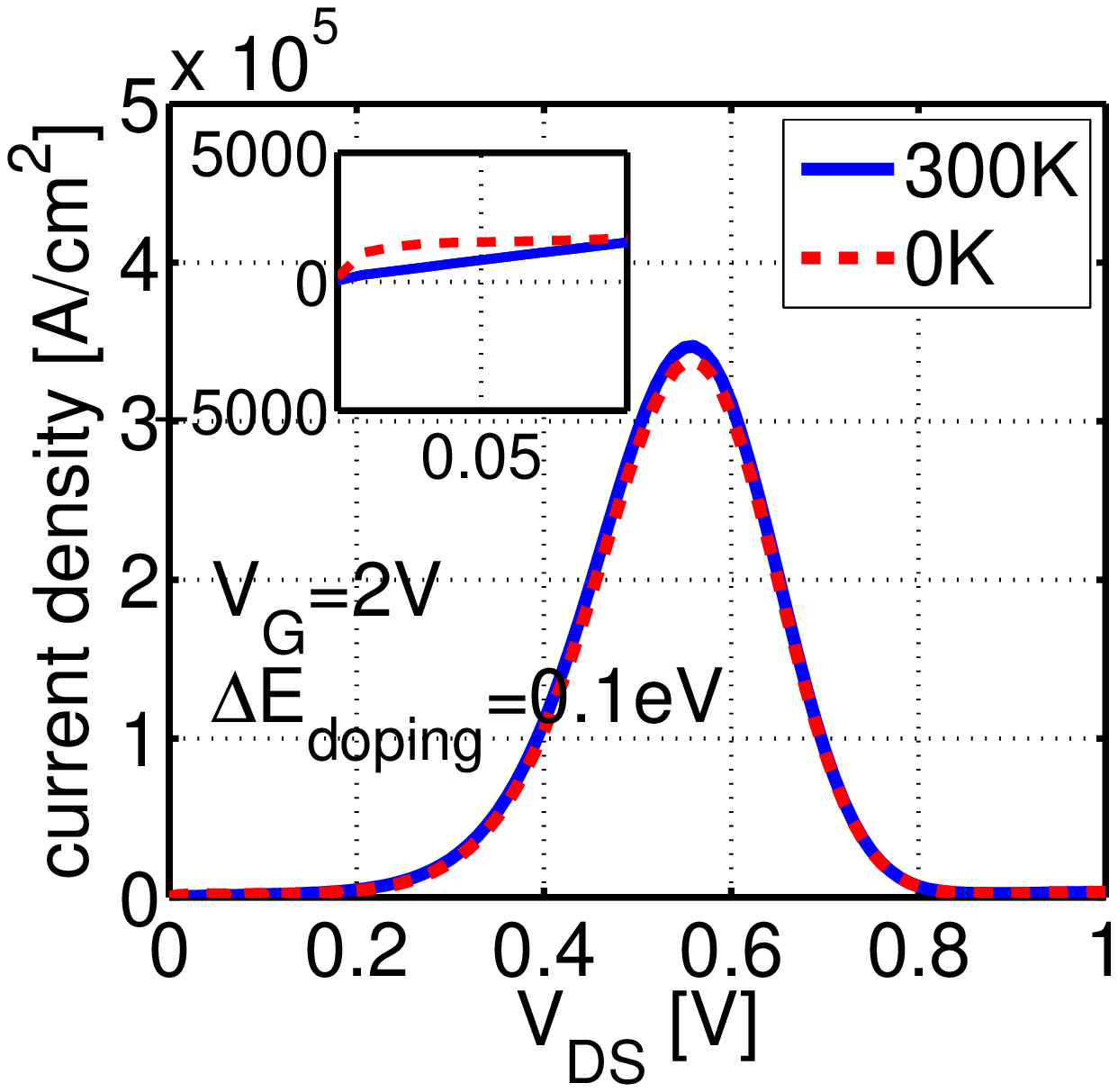}

}\subfloat[]{\includegraphics[width=0.22\textwidth]{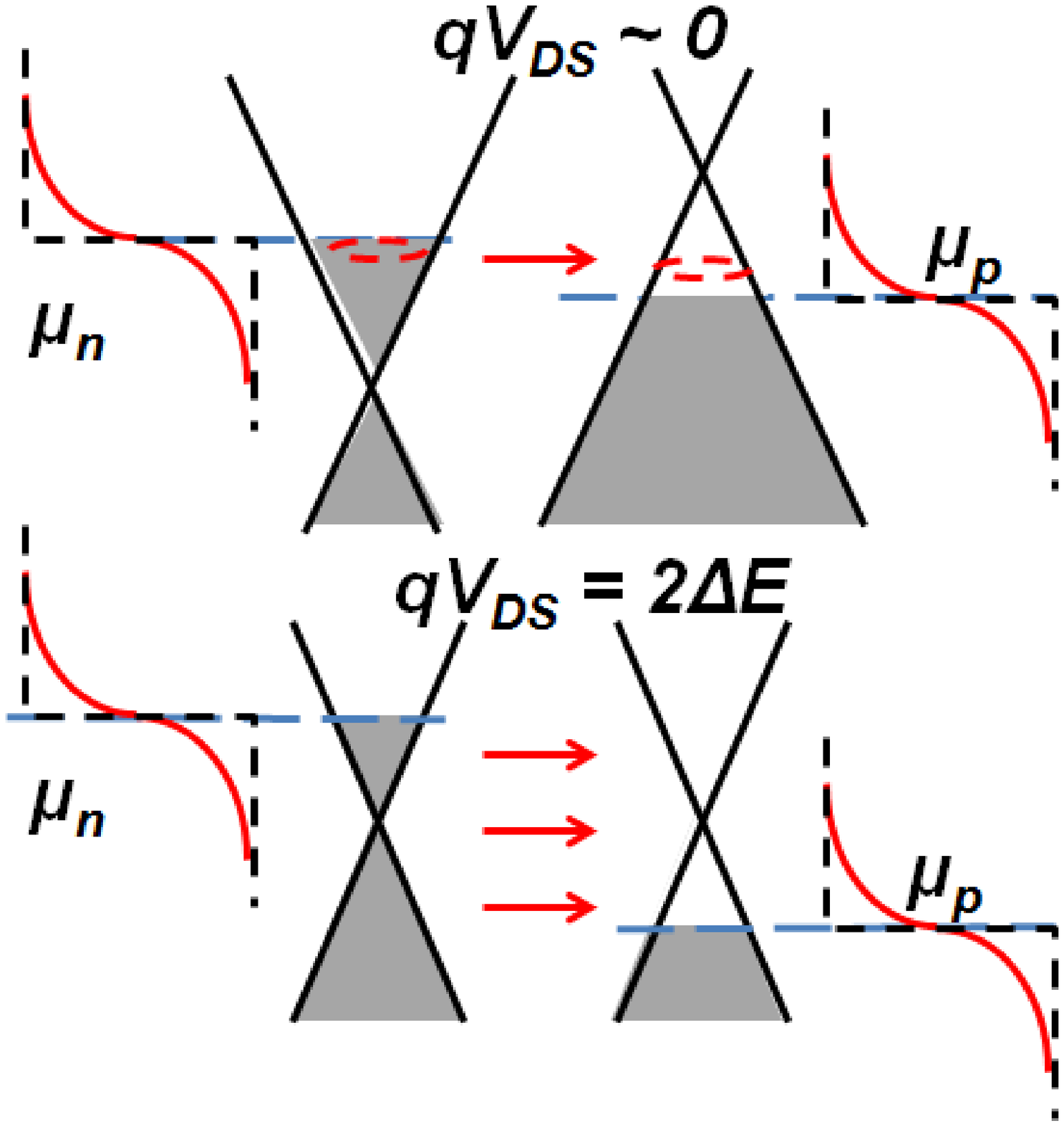}

}
\par\end{centering}

\centering{}\caption{ (a) The comparison of $I_{D}-V_{DS}$
characteristics at $T=300$K and $T=0$K. (b) the sketch of the band
diagram and Fermi Dirac distribution at $T=300$K and $T=0$K. The
inset of (a) shows the current density near $V_{DS}=0$}
\label{fig:fig5}
\end{figure}

Because tunneling is the main current transport mechanism, the
$I_{D}-V_{DS}$ curve is quite insensitive to temperature as shown in
Fig.~\ref{fig:fig5}.  But since the Fermi Dirac distribution smears
out state occupancy at a finite temperature, slight differences can
still be observed between $T=300$K and $T=0$K. At low $V_{DS}$, the
transport energy window (between the quasi-Fermi levels $\mu_{n}$
and $\mu_{p}$) is small (Fig.~\ref{fig:fig5} (b)). Then, the Fermi
distribution smearing reduces the carrier density at higher
temperature and the current decreases. The increase of the resonant
peak current at room temperature is because the Fermi distribution
tail extends to higher energy with more states. When the Dirac
points are aligned, states at all energies conserve lateral momentum
upon tunneling, and thus are allowed.

\begin{figure}[t]
\centering{}\includegraphics[width=0.4\textwidth]{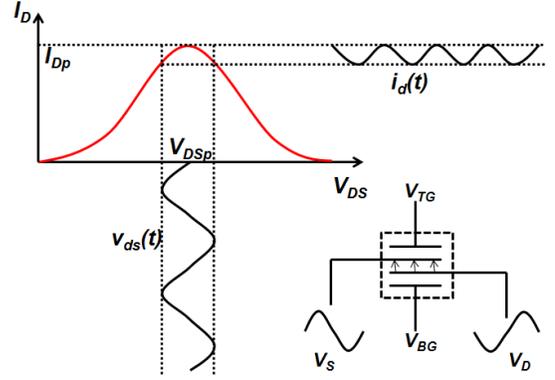} \caption{
The nonlinear resonant current is highly symmetric. When a DC
voltage $V_{DSp}$ biased at the current peak is superimposed with an
ac signal $v_{ds}(t)$, the frequency of the output current $i(t)$
will be doubled.} \label{fig:fig6}
\end{figure}

We note that the device current has symmetric resonances in both the
$I_{D}-V_{DS}$ and $I_{D}-V_{G}$ scans.  This is quite unlike what
happens in a single-layer graphene FET where the `ambipolar' nature
manifests itself primarily in the gate bias sweep \cite{Wang_10a}
\cite{Wang_10b}. The nonlinear symmetric resonant $I_{D}-V_{DS}$
behavior can be used for purpose of frequency multiplication
(Fig.~\ref{fig:fig6}).  If a DC voltage bias at the current peak
$V_{DSp}$ is superimposed with an ac signal, the frequency of the
output current will be doubled.  We can use equation \eqref{eq:Id3}
to calculate the output ac current signal. Assuming that
$V_{DS}=V_{DSp}+v_{ds}e^{i\omega t}$, the oscillatory part of the
current is:

\begin{equation}  \label{eq:peak}
I=\frac{1.6}{\sqrt{2\pi}}G_{1}\frac{L\Delta
E^{2}(2u_{11}^{4}+u_{12}^{4})}{u_{12}^{4}q\hbar
v_{F}}\exp(-\frac{A}{4\pi}[\frac{v_{ds}e^{i\omega t}}{\hbar
v_{F}}]^{2}).
\end{equation}
To find out the higher order harmonics, we ignore the constant
prefactor:

\begin{equation} \label{eq:peak2}
I\varpropto\exp(-\frac{A}{4\pi}[\frac{v_{ds}e^{i\omega t}}{\hbar
v_{F}}]^{2}).
\end{equation}
This expression can be further extended as:
\begin{equation} \label{eq:peak_series}
I\varpropto1-C_{1}\exp(2j\omega t)+\frac{C_{1}^{2}}{2!}\exp(4j\omega
t)-\frac{C_{1}^{3}}{3!}\exp(8j\omega t)+\cdots,
\end{equation}
where $C_{1}=\frac{A}{4\pi(\hbar v_{F})^{2}}v_{ds}^{2}$.  In
equation \eqref{eq:peak_series}, only even higher harmonics occur.

The SymFET is expected to be intrinsically fast since it relies
entirely on tunneling. The extrinsic performance with parasitics can
be analyzed same as for any high speed device and will not be
covered in this paper. High-frequency digital operation and a host
of analog applications such as frequency multiplication are thus
possible by exploiting the symmetry of the bandstructure of 2D
graphene.

As explained in \cite{Freenstra_12}, the greatest amount of
nonlinearity in the $I-V$ characteristics is achieved with nearly
perfect rotational orientation of graphene layers.  This presents a
significant challenge in fabrication of such devices based on the
layer transfer technology.  But epitaxial growth of graphene, BN
\cite{Nagashima_95} or other 2D materials provides a choice to
overcome this problem.  We note that due to the many-particle nature
of the excitonic condensate, the BiSFET is expected to be
insensitive to rotational misalignment \cite{Banerjee_09}.  This is
similar to the robustness of superconductivity to defects.  However,
the single-particle tunneling nature also makes the SymFET robust to
certain quantities to which the BiSFET is sensitive.  As discussed
earlier, the SymFET is robust to temperature.  Another advantage is
the robustness of single-particle tunneling - this is an intrinsic
advantage for the SymFET.  Although the tunneling current will vary
with the tunneling insulator thickness and its dielectric constant,
the regular single-particle tunneling behavior will survive in all
temperature and thickness values.  Thus, unlike the BiSFET's
sensitivity to thickness variations, the SymFET behavior is robust
to thickness and dielectric constant variations. Both thickness
variations and inelastic scattering processes can be significantly
suppressed by using 2D crystal `insulators' such as BN or MoS$_2$
between the graphene layers, preferably in a rotationally aligned
structure.

In summary, we have presented an analytical model to calculate the
channel doping potential and current-voltage characteristics in a
novel electronic device structure, the SymFET. The current in a
SymFET flows by tunneling from one graphene layer to the other.  The
current is insensitive to temperature.  The resonant current peak is
controlled by chemical doping and applied gate bias.  The on/off
ratio increases with graphene coherence length and doping.  The
symmetric resonant peak is a good candidate for high-speed analog
applications, of which frequency multiplication is an example.  The
resonant peak behavior can also be the framework for new digital
architectures that consume much lower power than the current state
of the art electronic switches.

\textbf{Acknowledgements} This work was supported by the
Semiconductor Research Corporation's Nanoelectronics Research
Initiative, the National Institute of Standards and Technology
through the Midwest Institute for Nanoelectronics Discovery (MIND),
and the National Science Foundation.

\end{document}